\documentclass[a4paper,10pt]{article}
\usepackage[utf8]{inputenc}

\usepackage{float}
\usepackage{graphicx}
\usepackage{amsmath}
\usepackage{amssymb}
\usepackage[round]{natbib}
\usepackage{amsfonts}

\newcommand\Rey{\mbox{\textit{Re}}}  

\newsavebox{\astrutbox}
\sbox{\astrutbox}{\rule[-5pt]{0pt}{20pt}}

\newcommand{\lsim}{\mathrel{\hbox{\rlap{\lower.55ex\hbox{$\sim$}} \kern-.3em 
 \raise.4ex \hbox{$<$}}}}

\newcommand{\gsim}{\mathrel{\hbox{\rlap{\lower.55ex\hbox{$\sim$}} \kern-.3em 
 \raise.4ex \hbox{$>$}}}}

\renewcommand{\v}[1]{\boldsymbol{#1}}

\DeclareGraphicsExtensions{.eps,pdf}
\graphicspath{{figures/}}

\title{On the forced flow around a flapping foil.}

\author{F. Mandujano and C. M\'alaga.}

\begin{document}

\maketitle

\begin{abstract}

The two dimensional incompressible viscous flow past a 
flapping foil immersed in a uniform stream is studied numerically. Numerical simulations were 
performed using a Lattice-Boltzmann model for moderate Reynolds numbers.
The computation of 
the hydrodynamic force on the foil is related to the the 
wake structure. In particular, when the foil's centre of mass is fixed in space, numerical 
results suggest a relation between drag coefficient behaviour and the flapping frequency which 
determines the transition from the von K\'arm\'an (vKm) to the 
inverted von K\'arm\'an wake. Beyond the inverted vKm transition 
the foil was released. Upstream swimming was observed at high 
enough flapping frequencies. Computed hydrodynamic forces suggest the propulsion mechanism
for the swimming foil. 


\end{abstract}



\section{Introduction}
\label{sec:int}

The flow around a flapping foil has been studied in a variety of
conditions. Foils on a stream, flapping as a result of the 
hydrodynamic forces acting on them, have been studied in 
connection to energy extraction processes ~\citep{jwu2015,tywu72}. The wake of foils
with an imposed flapping and translational motion, and its 
relation with the reacting forces on them has deserved attention 
due to its relation to flying and swimming~\citep{anderson,dong,triantafyllou,nttaylor}. 
Little has been done in the case of an imposed flapping on a foil 
free to swim, although work can be found related to 
other type of swimmers at moderate Reynolds numbers~\citep{kern,dabiri,lauga}. 
Here we study numerically a 
two-dimensional foil with an imposed flapping motion, immersed in 
a stream, that can be released and become free of translational 
motion. Both phenomena were studied and compared for moderate Reynolds numbers, 
where inertia and viscous forces are important. 


We used the lattice Boltzmann model (LBM) of~\citet{He98}, with 
a procedure to include immersed moving boundaries of arbitrary 
shape subject to hydrodynamic forces~\citep{mandujano08}. The LBM 
is an efficient algorithm to approximate solutions to the Navier-
Stokes equations when implemented on massively parallel 
architectures, where it can handle simulations of high resolution 
and large computational domains~\citep{clausen}. For the problem in hand the 
LBM seemed a good option, since wakes behind objects may show 
interesting behaviour far away behind the object~\citep{vorobieff}, and so a large domain relative to the size of the 
foil was desirable, along with enough resolution to observe 
vortex formation and evolution.



The article is divided as follows, in section \ref{sec:stat} the 
general problem is presented and the numerical method used is 
described briefly in section \ref{sec:lbe}. In section 
\ref{sec:val} the results are presented and discussed. Conclusions
are summarised in in section \ref{sec:conc}.


\section{Statement of the problem}
\label{sec:stat}
\begin{figure}
  \begin{center}
   \includegraphics[width=0.7\textwidth]{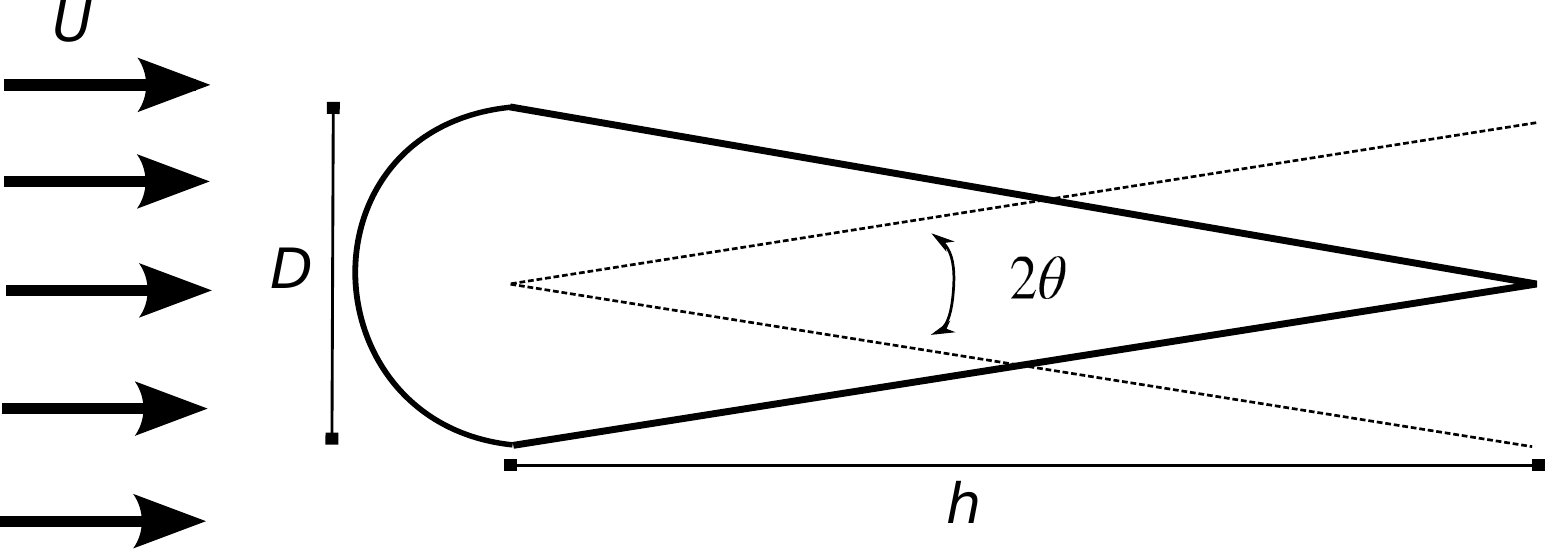} 
  \end{center}
  \caption{\label{fig:prob} Schematic illustration of the flow around a flapping foil.}
 \end{figure}

Consider an unbounded two dimensional and incompressible flow of a viscous fluid around a 
rigid flapping foil made by a semicircle, with diameter $D$, intersected with an isosceles 
triangle of high $h$, as shown in figure \ref{fig:prob}.
Two cases are considered, a foil with a point fixed in space and a foil free to move. In both 
cases the flapping motion consist of 
an oscillation with respect to the centre of the semicircle, which is fixed in space in one 
case and free in the other. To simplify 
the calculations, it is assumed that the foil mass density distribution, $\rho_s$, is such 
that its centre of mass is at the centre of the semicircle.

The governing equations for the fluid, with mass density $\rho_f$ and kinematic viscosity 
$\nu$, are the Navier-Stokes equations given by
\begin{eqnarray}
\nabla \cdot \v{u} &=& 0,\label{cont} \\
\frac{\partial \v{u}}{\partial t} + \v{u}\cdot \nabla \v{u} &=& -\frac{1}{\rho_f} \nabla P + \nu \nabla^2 \v{u}, \label{mom}
\end{eqnarray} 
where $\v{u}(\v{r},t)$ and $P(\v{r},t)$ are the velocity and pressure fields, respectively. 
Consider no-slip condition at the foil's surface. Constant pressure and a uniform and 
horizontal velocity field $\v{U}$ are imposed far from the foil (see figure \ref{fig:prob}).
 
The rigid foil is forced to rotate around the centre of the semicircle (also its centre of mass) so that the angle of its axis of 
symmetry with respect to the horizontal coordinate evolve with $\theta(t) = \theta_0 \sin \omega t$ (see Figure \ref{fig:prob}). The 
velocity of a point at the foil's surface is $\v{v}_s({\bf r}_s,t)=\v{V}(t) + \dot{\theta}(t) (\v{r}_s-\v{R})$, where $\v{R}$ is the 
position of the centre of mass, $\v{V}$ its velocity and $\v{r}_s$ is the position of a point at the foil's surface. Therefore, 
the no-slip boundary condition takes the form $\v{u}(\v{r}_s,t) = \v{v}_s(\v{r}_s,t)$. 
In the case of the free flapping foil, the motion of its centre of mass is obtained by solving $m d\v{V}/dt = \v{F}$, where $m$ is
the mass of the foil. The hydrodynamical force $\v{F}$ acting on the foil, is obtained directly from the computed flow fields.


To work with non-dimensional variables we choose $D$ as the characteristic length, $U$ as the characteristic velocity and the fluid 
density $\rho_f$. With this choice of scaling the flow regime is determined by five dimensionless parameters: the Reynolds number 
$\Rey = UD/\nu  $, the Strouhal number $ St = \omega D / 2\pi U $ the dimensionless amplitude $A = 2 \theta_0 h/D$, the ratio 
between the chord of the foil and the circle's diameter $ C = 1/2+h/D $, and the fluid-foil's mass density ratio $\mu = 
\rho_f/\rho_s$ which is set to unity to work with a non-buoyant foil. The hydrodynamic force is scaled with $\rho_f U^2 D/2$. We 
define the drag $C_D$ and lift $C_L$ coefficients as the horizontal and vertical components of $2\v{F}/\rho_f U^2 D$, respectively.
Notice that negative values of $C_D$ correspond to a thrust dominated hydrodynamic force.

Solutions to equations (\ref{mom}) and (\ref{cont}) for the pressure and velocity fields are approximated using a lattice-Boltzmann
model. The numerical procedures are described in the following section.
 
\section{The numerical scheme}
\label{sec:lbe}

To compute the flow around the flapping foil a two dimensional, nine neighbours (D2Q9) lattice-Boltzmann model was
used \citep{He98}. The proposed algorithm, that includes moving immersed boundaries, was validated in a previous work concerning a 
problem of a moving cylinder in a convective flow~\citep{mandujano08}. 

In this method space is discretised using a square lattice. Lattice spacing as well as time steps can be 
conveniently set to unity.
The state of the fluid at the node with vector position $\v{r}$ at time $t$, is described by the particle distribution function $f_k(\v{r},t)$ that evolves in time and space according to 
\begin{equation}
 f_k(\v{r}+\v{e}_k,t+1) = f_k(\v{r},t)-\dfrac{1}{\tau}%
 	\left[f_k(\v{r},t)- f_{k}^{(eq)}(\v{r},t)\right],\label{eq:lbed}
\end{equation}
where $\tau$ is a relaxation time related to the fluid kinematic viscosity $\nu = (\tau- 1/2)/ 3$. The distribution function $f_{k}^{(eq)}$ is given by
\begin{equation}
 f_{k}^{(eq)}(\v{r},t)= w_k\rho\left(1 + 3\v{e}_k\cdot\v{u}+
 \dfrac{9}{2}(\v{e}_k\cdot\v{u})^{2}-\frac{3}{2}\v{u}^2\right),\label{eq:feq}
\end{equation}
which corresponds to a discrete Maxwell distribution function for thermal equilibrium. In the above expressions, the macroscopic 
density and velocity fields are computed using
\[
 \rho(\v{r},t) = \sum_{k}f_k(\v{r},t) \quad \textmd{and} \quad
 \rho {\v{u}}(\v{r},t) = \sum_{k}\v{e}_k f_k(\v{r},t).
\]
The microscopic set of velocities $\v{e}_k$ is given by
\begin{align*}
 \v{e}_0= (0,0),\quad 
 \v{e}_k=\left(\cos(\pi(k-1)/2),\sin(\pi(k-1)/2)\right) \ \ \textmd{for} \ \ k=1,\dots,4,\\
 \v{e}_k=\sqrt{2}\left(\cos(\pi(k-9/2)/2),\sin(\pi(k-9/2)/2)\right) \ \ \textmd{for} \ \ 
 k=5,\dots,8,
\end{align*}
where $w_0=4/9$, $w_k=1/9$ for $k=1,\dots,4$ and $w_k=1/36$ for $k=5,\dots,8$. Notice that, with the choice of microscopic velocities $\v{e}_k$, expression (\ref{eq:lbed}) is always evaluated at lattice points. It is well known that the above procedure approximates solutions to the Navier-Stokes equations in the limit of small Mach numbers \citep{He98}.

Equation (\ref{eq:lbed}) provide an explicit algorithm for updating all the distribution functions $f_k$ at a given node in the lattice, as long as its 8 nearest neighbouring nodes are inside the fluid domain. For nodes adjacent to a solid wall, the distribution functions coming from neighbouring nodes outside the fluid domain must be provided as a boundary condition for the method. We choose to adopt the set of boundary conditions proposed by \citet{guo02a} for curved rigid walls.

The force and torque acting on the body are computed using the momentum-exchange method of 
\citet{mei02}. For the free flapping foil, as 
rotational motion is imposed, only the force is used to compute the foils translational motion using a 
forward Euler integration in time at each time step. This is a particular case of the scheme described in 
\citet{mandujano08} for 
the motion of a free particle in a convective flow using the lattice-Boltzmann model. 

The computational domain was a rectangular lattice of $10000 \times 3000$ nodes for the fixed foil, and of $12000 \times 5000$ nodes
for the free foil. The foil had a length $h$ of 280 to 600 nodes, a width $D$ of 80 nodes (about a $1/30$ of the width and length of the domain), and was placed at 3000 nodes from the left boundary of the domain ($5$ foil lengths). To simulate conditions far from the foil, the velocity was set to $\v{U}$ at the left boundary following the procedure of \cite{guo02a}. On the rest of the boundaries, the normal components of the velocity gradients were set to zero setting the unknown velocity at the boundary equal to that of the adjacent node normal to the wall. 

The numerical scheme was implemented to run in parallel in GPU's due to the large number of nodes involved in the simulations. 
Typically, two days are needed to obtain a periodic flow ($3\times 10^5$ time steps) of a free foil swimming upstream running on an Nvidia\textregistered Tesla K40 processor.


\section{Results}
\label{sec:val}

The foil profile geometry was chosen to compare with experiments performed by \cite{bohr} in a soap films and 
by \cite{godoy} in a wind tunnel. Numerical results confirm 
the relation between the drag-trust transition and the behaviour of the von Karman (vKm) vortex street.  

\subsection{A fixed flapping foil}

Simulations start with the foil at rest facing a uniform flow at the left boundary to allow for a wake to form 
before flapping. After $5\times 10^4$ time steps, the foil starts to flap increasing its amplitude 
exponentially in time until it reaches $A$. The simulations were performed for a series of values of $St \in [0.1,0.6]$,
$A\in [0.5,3]$, $Re \in [100,255]$, and $C$ took values $4$ and $6$.
\begin{figure}
  \begin{center}
  \includegraphics[width=1.0\textwidth]{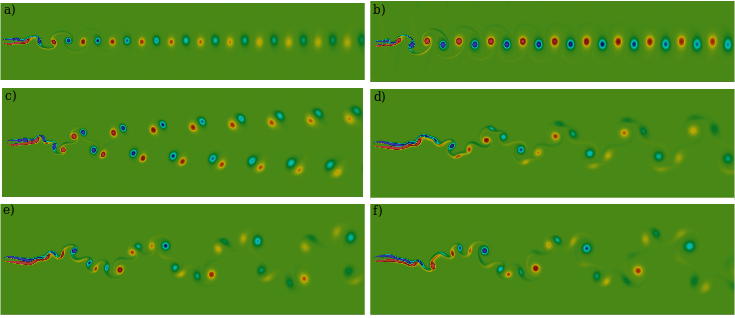} 
  \end{center}
  \caption{\label{fig:forces1} Vorticity distribution for a fixed flapping foil for $\Rey=220$. Colours represent positive (red) and negative (blue) vorticity. a) vKm wake for $ St=0.12 $ and 
  $A=0.98 $. b) inverted vKm wake for $St=0.12$ and $A=2$. c) 2P wake for $St=0.085$ and $A=1.4$. d) 2P+2S wake for $St=0.05$ and 
  $A=1.4$. e) 4P wake for $St=0.035$ and $A=1.34$. f) 4P+2S wake for $St=0.025$ and $A=1.2$. Figures show the region of interest within the computational domain.}
 \end{figure}

Following \cite{bohr}, in the first set of simulations $\Rey=220$ and $C=6$. The simulations performed show 
the formation of patterns in approximately the same parameter region as reported there (see Figure \ref{fig:forces1}); 
transitions between vKm and inverted vKm wake, formation of 2P, 4P, 2P+2S, 4P+2S 
wakes (following the nomenclature of \citet{williamson}) and the transitions between them.  
Quantitative comparison can not be expected as film flow is not a two dimensional phenomena.

\begin{figure}
  \begin{center}
   \includegraphics[width=1.08\textwidth]{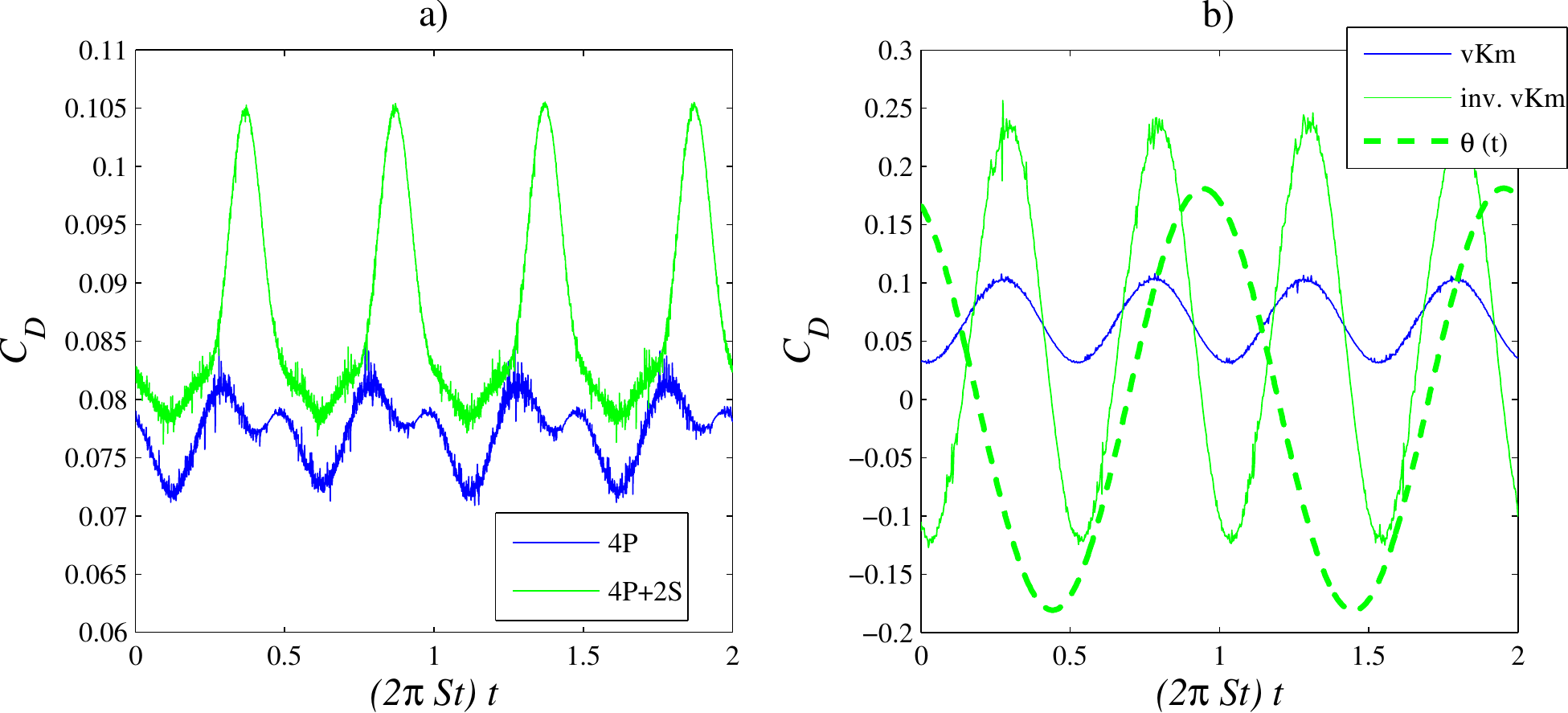} 
  \end{center}
  \caption{\label{fig:drag} Drag coefficient as function of time for some of the vorticity distributions of Figure 
  \ref{fig:forces1}. a) 4P wake for $St=0.035$ and $A=1.34$ and 4P+2S wake for $St=0.025$ and $A=1.2$ , b) vKm wake for $St=0.12$ and $A=0.98$ and inverted vKm wake for $St=0.12$ and $A=2$. The angle of attack $\theta (t)$ of the foil is shown in (b) for the inverted vKm case.}
 \end{figure}

The computed hydrodynamical force on the foil showed a periodic dependence on time (see Figure \ref{fig:drag}). 
When the flow has 2S or 2P wakes (Figures \ref{fig:forces1}-(a)-(c)), the lift coefficient $C_L$ oscillates with the flapping 
frequency, corresponding to a Fourier mode $n=1$. The drag coefficient $C_D$ oscillates with twice the flapping frequency and has
modes $n=0$ and $n=2$. Notice that $C_D$ minima (maximum thrust) is obtained after extremal values of the angle of
attack $\theta$ are reached.

The flapping frequency sets the hydrodynamic force frequencies and higher harmonics. For 
patterns with combinations of 2P, 4P and 2S, Fourier coefficients corresponding to higher 
harmonics become important. In Figure \ref{fig:forces1}-(d)-(f), the lift and torque 
coefficients are composed by the first few odd modes while $C_D$ modes are even 
(see Figure \ref{fig:drag}-(a)). In the combination 2P+2S, $C_D$ have $n=2$ and 4 wile $C_L$ 
is composed by $n=1$ and $3$ modes. The 4P wake includes modes $n=6$ and $n=7$ and the 4P+2S wake has $n=8$ 
and $n=9$ modes. $C_D$ always showed even modes while $C_L$ modes were all odd. There is a strong 
horizontal push with every stroke and higher harmonics are produced at low flapping 
frequencies where many vortices are shed during each stroke.

When $St=0.12$ and $A \in [0.98,3]$ there is a transition between the vKm and the inverted vKm wake, known to be relate with the 
drag-thrust transition~\citep{karman,bohr,godoy}. Our LBM simulations confirm that there is an inversion of the vKm wake in this 
interval accompanied by a mostly sinusoidal $C_D$ with a minimum value that crosses zero at the transition, while
retaining a positive mean value (see Figure \ref{fig:drag}-(b)). 
 
\begin{figure}
  \begin{center}
  \includegraphics[width=1.05\textwidth]{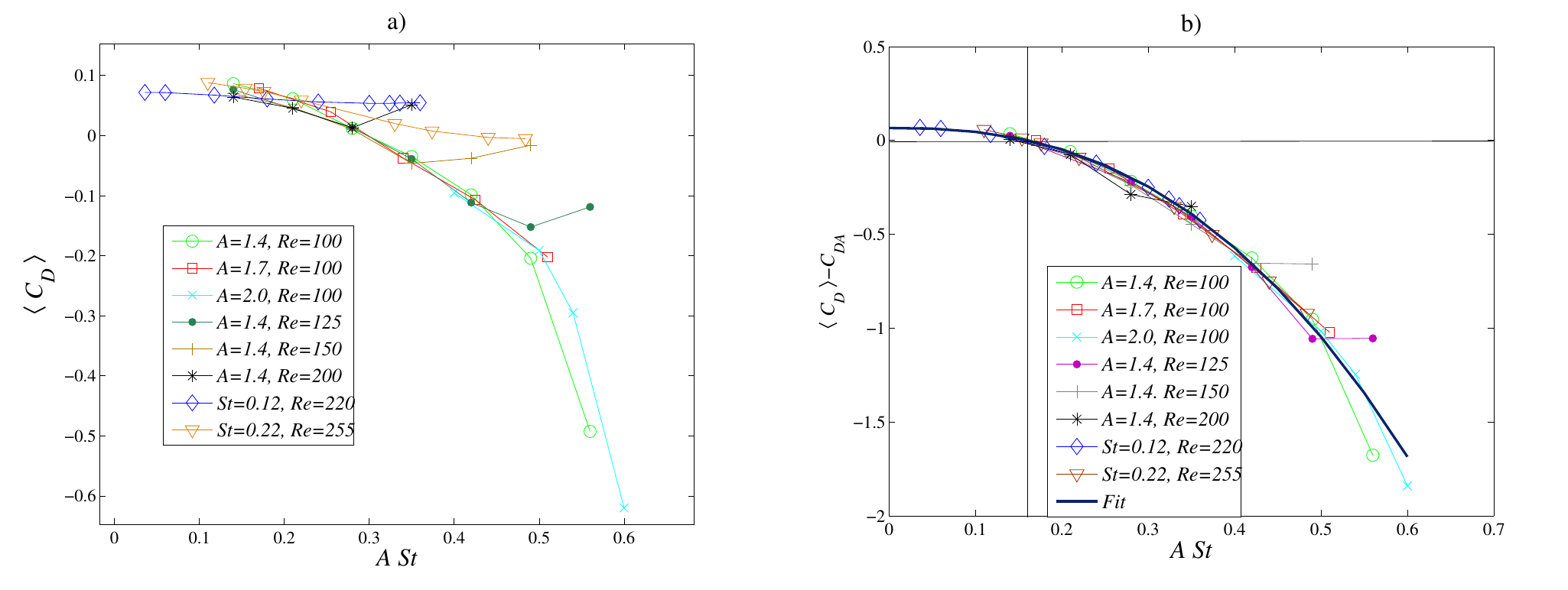}   
  \end{center}
  \caption{a) Mean drag coefficient $\langle C_D \rangle$ and b) $\langle C_D \rangle - C_{DA}$ as functions of the $StA$. 
  The lines connecting  points mean that either $A$ or $St$ were kept constant as marked. The last two experiments correspond to the parameters used in \cite{bohr} and  \cite{godoy}, respectively. $C=4$ when $Re=255$ and $C=6$ in the rest of the cases. \label{advscd} }
 \end{figure}

In Figure \ref{advscd} it is shown the mean drag coefficient $\langle C_D \rangle$ and the difference between $\langle C_D \rangle$ 
and the norm of its Fourier 
decomposition $C_{DA}$, as a function of the product $StA$. For a sinusoidal $C_D$, $\langle C_D \rangle 
- C_{DA}$ is the minimum value of $C_D$.

Figure \ref{advscd}(a) shows that the mean drag decreases monotonically and becomes negative, as $StA$ increases for the
cases where $Re < 150$. Numerical experiments with higher $Re$ show that $\langle C_D \rangle$ decreases and can be negative
but starts growing beyond certain value of $StA$. These observations suggest that for a given imposed flow $U$, to which $Re$ 
is related, there is a critical flapping frequency beyond which the mean drag starts growing. We could not confirm this idea for 
the cases where $Re < 150$ because the numerical model became unstable for high flapping frequencies.
\begin{figure}
  \begin{center}
   \includegraphics[width=1.0\textwidth]{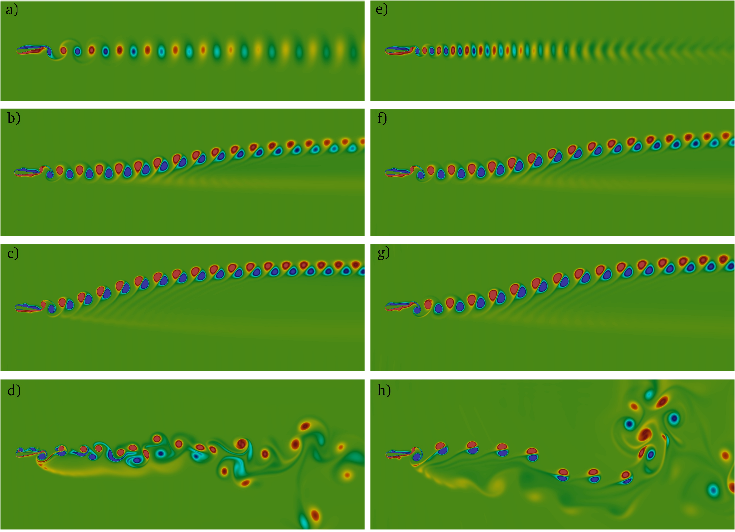} 
  \end{center}
  \caption{\label{fig:st0p12Advar} Evolution of the wake for $Re=100$ and $ C=6 $. a) $St A=0.21$, b) $St A=0.42$,
  c) $St A= 0.49 $, d) $St A=0.56$, e) $St A=0.21$, f) $St A= 0.42$, g) $St A= 0.51$ and h) $St A= 0.6$. In the left column the 
  dimensionless amplitude was keep fixed at $A=1.4$ and, in the right column, the Strouhal number was keep fixed at $S=0.3$. Figures show the region of interest within the computational domain. Colours represent positive (red) and negative (blue) vorticity.}
 \end{figure}
Figure \ref{advscd}(b) shows that the minimum of $C_D$ crosses zero at $StA\sim 0.16$ 
regardless of the values of $Re$ and $C$ used.
The numerical experiments show that below this value of $StA$ vortices at the wake 
corresponds to the vKm wake type, but can be combinations of 2S and 2P patterns when $St$ is 
small. The inverted vKm is found when $StA>0.16$ (see figures \ref{fig:forces1}(b), 
\ref{fig:st0p12Advar}(a) and \ref{fig:st0p12Advar}(e)). Figure 
\ref{advscd}(b) shows that all cases collapse approximately on a single curve given by the fit
\begin{equation}
    \langle C_D \rangle - C_{DA} = -6.25 (StA) ^{2.5}+0.066.
    \label{fit}
\end{equation}

For $\Rey=220$ and $C=6$, the wake remains symmetric with a positive value of $\langle C_D \rangle$
for all values of $StA$ explored by ~\citet{bohr}. The evolution of the wake as 
function of $A$ and $St$, starting form a vKm wake, is shown in figure \ref{fig:st0p12Advar} 
for $Re = 100$. Beyond the inverted vKm transition, for $StA > 0.28$, the wake is deflected 
from the horizontal centre line and $\langle C_D \rangle$ becomes negative. The results shown that the distance between the tail of the 
foil and the point of deflection seems to decrease as $StA$ increases, in agreement with the numerical simulations made by \cite{deng} (see figures \ref{fig:st0p12Advar}(b), (c), (f) and (g)). For $Re=255$, the 
performed simulations shown that the wake deflects when $St A > 0.23$, in good agreement with 
the 3D experimental observations of \citet{godoy} and the numerical simulations of \cite{guo-yi12} and \cite{deng}. 

The direction of deflection is related with the initial condition, 
and a downwards deflection was found when changing the sign of the angle $\theta_0\rightarrow 
-\theta_0$. With the set of parameters used by ~\citet{godoy}, the simulations showed that the 
position and angle of deflection can change with time after several vortex shedding periods of 
observation, this long time behaviour has been observed by \citet{lewin2003}. 

The results with $\Rey=100$ at values of $St A \sim 0.5$ show that the wake becomes unstable 
(see figures \ref{fig:st0p12Advar} (d) and (h)). The vorticity produced by the foil is still 
periodic but highly asymmetric, with $C_L$ showing only even modes while $C_D$ showing 
contribution from both even and odd modes.

\subsection{A free flapping foil}

The simulations for the free foil were made using similar initial conditions as in the fixed 
case. After $5\times 10^4$ iterations, where the 
foil was fixed without flapping, the foils is released and starts flapping with the same 
exponential increase in amplitude as the 
fixed case. The values of $StA \in [0.1,0.5]$ were chosen in correspondence to patterns beyond 
the transition to the inverted vKm wake for the fixed case.

\begin{figure}
  \begin{center}
   \includegraphics[width=1.\textwidth]{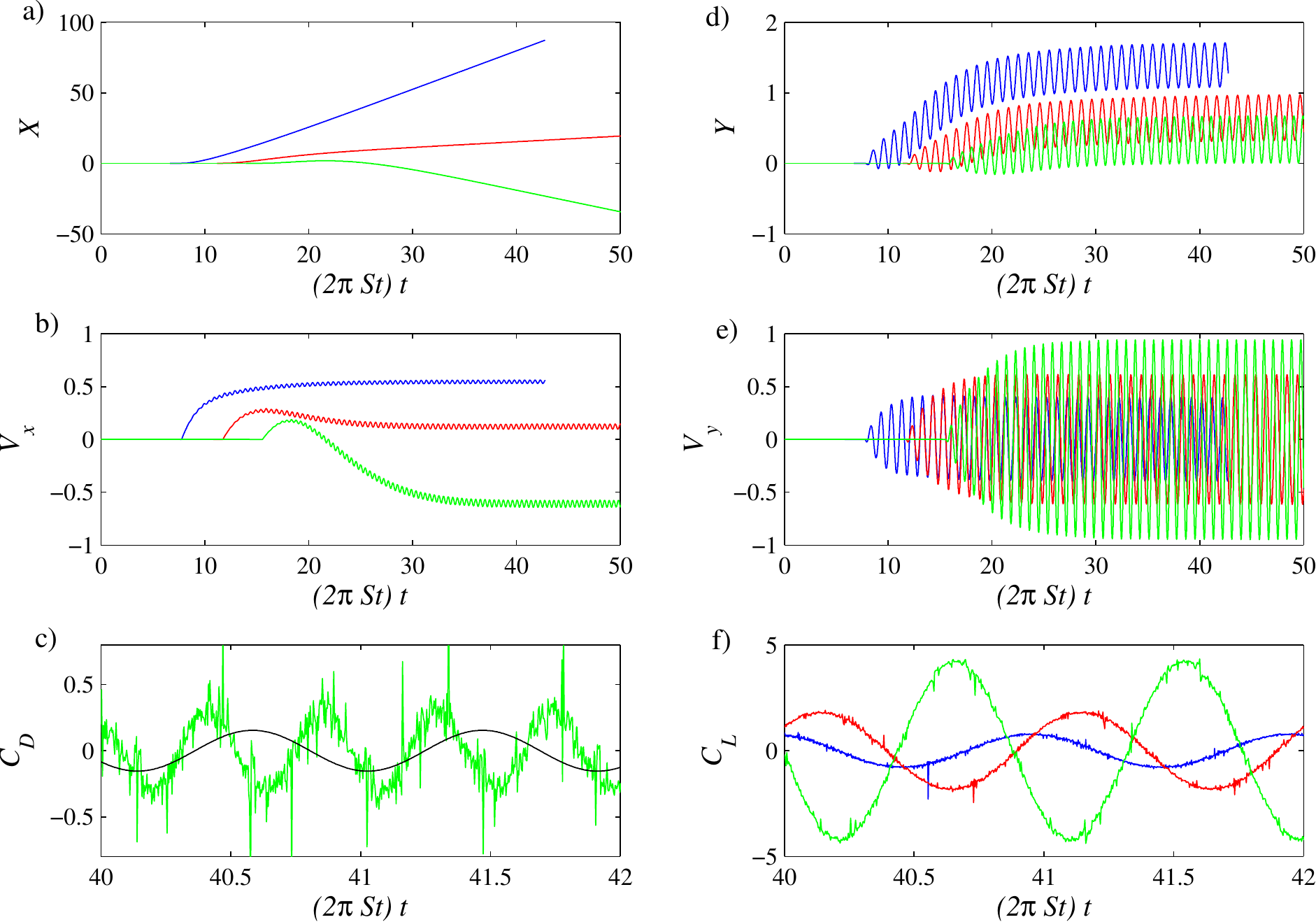}
  \end{center}
  \caption{Dynamics of the centre of mass of the foil for $A=1.7$, a) horizontal position 
  $X(t)$, b) 
  horizontal velocity $V_x(t)$, c) Drag coefficient $C_D(t)$, d) vertical position $Y(t)$ , e) 
  vertical velocity $V_y(t)$ and f) Lift coefficient $C_L(t)$. (Blue) $St=0.2$, (red) $St=0.3$ 
  and (green) $St=0.45$ . The black line in (c) shows the angle of attack $\theta (t)$. }
\label{fig:free1}
\end{figure}

A stationary symmetric flow without vortex shedding was observed before the foil was released 
in all cases. After released, a transient time was observed where the hydrodynamic force 
changed from a constant value to a periodic function of time. During this time interval, the 
centre of mass accelerated until it reached an almost uniform motion in the horizontal 
direction with small transversal oscillations of the order of $D$ (see figures \ref{fig:free1}(a) and 
(d)). The centre of mass deviates vertically a small distance from the horizontal centre line during the transient, probably related to the deflection of the wake.

The centre of mass velocity is a periodic function of time, the horizontal component 
oscillates around a constant value that is a fraction of the velocity of the free stream 
($U=1$ in dimensionless units) with an amplitude that is negligible, hence the swimming 
velocity is practically constant  
as shown in figures \ref{fig:free1} (a) and (b). The observed amplitude of oscillation of 
$C_L$ is ten times bigger than that of $C_D$, which produces a more appreciable oscillation in the 
vertical direction (see Figures \ref{fig:free1} (d)-(f)).

As $St A$ is increased, $V_x$ diminishes until it becomes negative and the foil swims upstream. The mean drag becomes very small, 
$\langle C_D \rangle \sim 10^{-4}$ in all cases observed, which is consistent with an almost 
uniform motion observed at long times in other swimmers~\citep{lauga,kern}. Figure \ref{fig:free1} (c) shows that minimum values of 
$C_D$ (maximum thrust) are
reached little after extremal values of the angle of attack of the foil, suggesting that swimming is not produced by 
pushing fluid but by a suction mechanism
resemblant of that observed by \citet{dabiri} in efficient animal swimming.

\begin{figure}
  \begin{center}
   \includegraphics[width=1.0\textwidth]{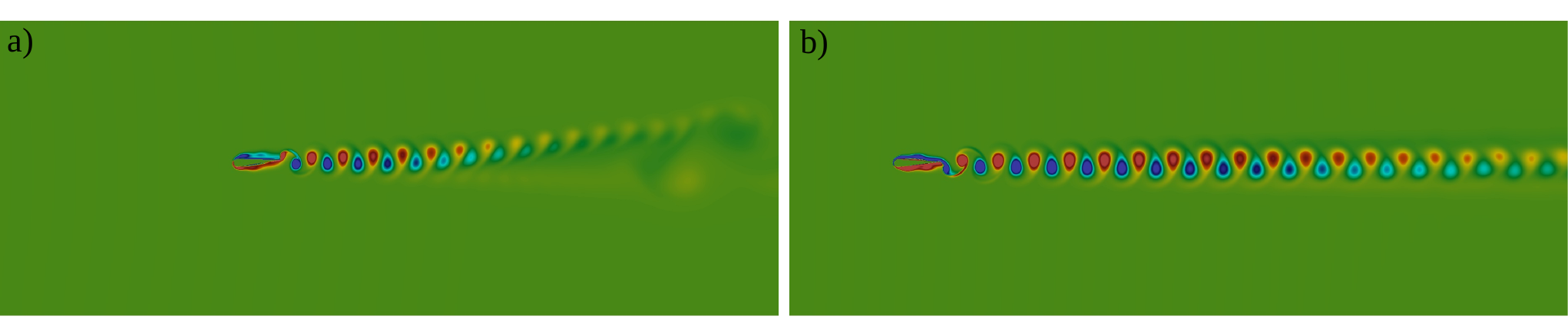}
  \end{center}
\caption{Vorticity distribution for a moving flapping foil for $C=6$, $A = 2.0$ and $Re = 100$. Colours represent positive (red) and negative (blue) vorticity. a) $ St = 0.2 $ and $V_x > 0 $. b) $ St = 0.3$ and $V_x < 0 $, the foil swims upstream. Figures show the complete computational domain. \label{fig:free} }
\end{figure}

\begin{figure}
  \begin{center}
   \includegraphics[width=1.0\textwidth]{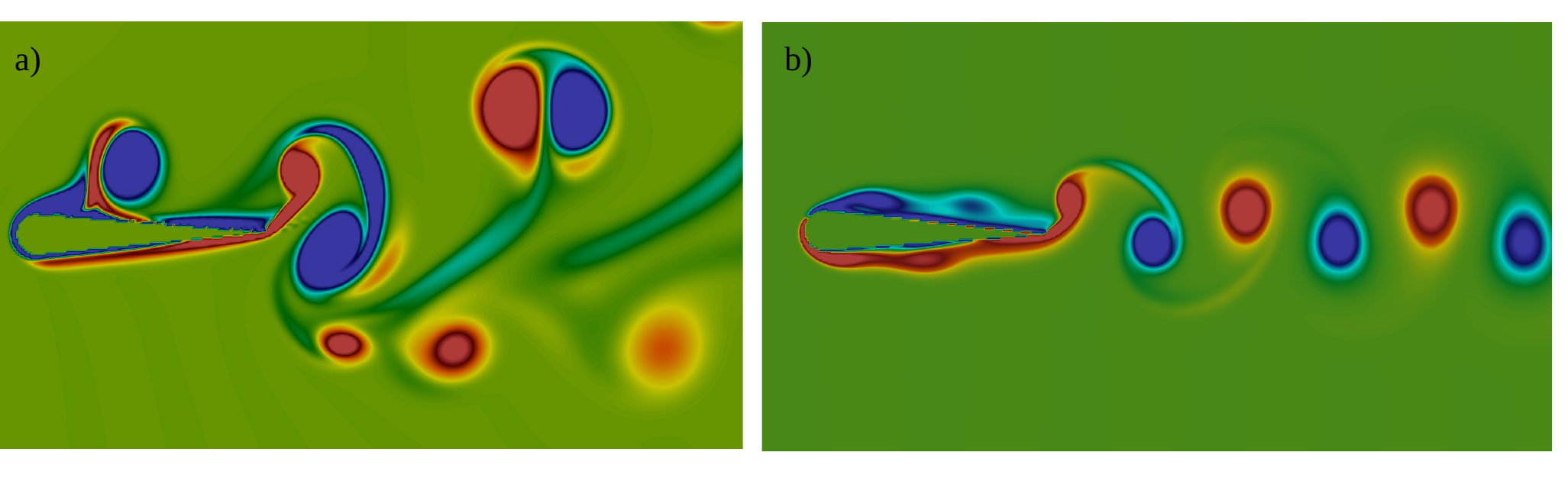}
  \end{center}
\caption{Close up view of the vorticity generated by the foil for $Re=100$, $St = 0.3$, and $A=2$. (a) Fixed foil of figure \ref{fig:st0p12Advar}-(h) that shows a P+S wake. (b) Free foil of figure \ref{fig:free}-(b) that shows an inverted vKm wake. \label{closeup} }   
\end{figure}

The wake behind the foil in this region corresponds to a inverted asymmetric vKm wake. 
When $V_x$ is positive (figure \ref{fig:free}-(a)) the angle of deflection of 
the wake seems to be smaller than those observed in the fixed foil cases of similar $StA$ values
(see figures \ref{fig:forces1} (b) and (f)). As $St A$ is 
increased and $V_x$ becomes negative (figure \ref{fig:free}-(b)), the deflection of the wake is swept away and becomes symmetric and much longer than the wakes of foils unable to swim upstream. The parameter values for the 
free foil swimming upstream correspond to the unstable wake for the fixed case (see figure \ref{fig:st0p12Advar}-(h)), a P+S mode as seen in figure \ref{closeup}(a). Figure \ref{closeup} shows how vortical structures at the foil's surface produce by fixed foil 
of unsteady wake are larger than those produced by the upstream swimming foil. Probably, the momentum produced by the foil, enough to swim upstream when released, competes with that of the stream to produce the instability.


\section{Conclusions}
\label{sec:conc}

In this work numerical results on the two-dimensional incompressible viscous flow around a flapping foil are presented. 
Both a fixed and a free flapping foil were considered. The flow field 
was computed using a lattice Boltzmann model that includes a procedure for moving internal 
boundaries with arbitrary geometry. Thus allowing the computation of the translational motion 
of a free foil due to hydrodynamic forces. Numerical results were compared with experiments on soap films \citep{bohr} 
and on wind tunnels \citep{godoy}. 

The numerical simulations show that the transition between the vKm wake and the inverted 
vKm wake on a fixed flapping foil is found when the flapping Strouhal number $StA \sim 
0.16$ and coincides with the minimum value of the horizontal hydrodynamic force on the foil crossing zero. When $St \lesssim 0.16$
the wake can be a 2S (vKm) or 2P. Above 
this value the wake corresponds to an inverted vKm that deflects as $StA$ increases and the mean horizontal
hydrodynamic force becomes negative. 
 
Whenever  $\langle C_D \rangle$ is a decreasing function of $StA$, there is a universal 
behaviour for all $Re$ and $C$ explored, with a power law fit given by equation (\ref{fit}). 
This result can be related with the narrow range of $A St$ where many mammals 
swims~\cite{nttaylor}. Results suggest that above a certain value of the flapping frequency 
the drag exerted on the foil starts growing probably producing a flapping foil unable to swim 
upstream if released.  

The cases studied for a flapping foil free of translational motion always reached an 
almost  uniform horizontal speed with a drag coefficient value practically zero. As 
flapping frequency increases the horizontal velocity becomes negative (upstream motion) and the deflected 
portion of the wake is swept away. Upstream swimming produces long, symmetric and 
inverted vKm wakes, and is observed beyond the vKm wake transition, when the fixed foil's wake becomes unstable.
Maximum thrust is observed close to maximal values of the foil's angle of attack, suggesting swimming through a suction 
mechanism.\\


Partial support from project UNAM-PAPIIT-IN115216 and IN115316 is acknowledged. Authors thank Dr. Eduardo Ramos and Dr. Ra\'ul Rechtman
for their support.



\end{document}